\documentclass{ws-procs9x6}
\newcommand{\be}{\begin{eqnarray}}
\newcommand{\ee}{\end{eqnarray}}
\newcommand{\bea}{\begin{eqnarray}}
\newcommand{\eea}{\end{eqnarray}}

\def\xbj{x_{\mbox{\tiny B}}}

\begin{document}

\title{Spin-Orbit Correlations}

\author{M. Burkardt$^*$}

\address{Physics Department, New Mexico State University,\\
Las Cruces, NM 88003, U.S.A.\\
$^*$E-mail: burkardt@nmsu.edu}

\begin{abstract}
We summarize the intuitive  connection between deformations
of parton distributions in impact parameter space and
single-spin asymmetries.
Lattice results for the $x^2$-moment of the twist-3 polarized 
parton distribution $g_2(x)$ are used to estimate the average
transverse force acting on the active quark in SIDIS in the
instant after being struck by the virtual photon.
\end{abstract}

\keywords{Style file; \LaTeX; Proceedings; World Scientific Publishing.}

\bodymatter
\section{Impact Parameter Dependent PDFs and SSAs}
\label{sec:IPDs}
The Fourier transform of the GPD 
$H_q(x,0,t)$ yields the  
distribution $q(x,{\bf b}_\perp)$ of 
unpolarized quarks, for an unpolarized target, in 
impact parameter space [\refcite{mb1}] 
\be
q(x,{\bf b}_\perp)= 
 \int \!\!\frac{d^2{\bf \Delta}_\perp}{(2\pi)^2}  
H_q(x,0,\!-{\bf \Delta}_\perp^2) \,e^{-i{\bf b_\perp} \cdot
{\bf \Delta}_\perp}, \label{eq:GPD}
\ee 
with ${\bf \Delta}_\perp = {\bf p}_\perp^\prime -{\bf p}_\perp$.
For a transversely polarized target (e.g. polarized in the
$+\hat{x}$-direction) the impact parameter dependent
PDF $q_{+\!\hat{x}}(x,{\bf b}_\perp)$ is
no longer axially symmetric and
the transverse deformation is described
by the gradient of the Fourier transform of the GPD $E_q(x,0,t)$
[\refcite{IJMPA}]
\bea
q_{+\!\hat{x}}(x,\!{\bf b_\perp}) 
&=& q(x,\!{\bf b_\perp})
- 
\frac{1}{2M} \frac{\partial}{\partial {b_y}} \!\int \!
\frac{d^2{\bf \Delta}_\perp}{(2\pi)^2}
E_q(x,0,\!-{\bf \Delta}_\perp^2)\,
e^{-i{\bf b}_\perp\cdot{\bf \Delta}_\perp}
\label{eq:deform}
\eea
$E_q(x,0,t)$ and hence the details of this deformation are not 
very well known, but its $x$-integral, the Pauli form factor
$F_2$, is. This allows to
relate the average transverse deformation resulting from Eq.
(\ref{eq:deform}) to the contribution from the
corresponding quark flavor to the anomalous magnetic moment.
This observation is important in understanding the sign of the
Sivers function.

In a target that is polarized transversely ({\it e.g.} vertically), 
the quarks in the target 
nucleon can exhibit a (left/right) asymmetry of the distribution 
$f_{q/p^\uparrow}(\xbj,{\bm k}_T)$ in their transverse 
momentum ${\bm k}_T$ [\refcite{sivers,trento}]
\be
f_{q/p^\uparrow}(\xbj,{\bm k}_T) = f_1^q(\xbj,k_T^2)
-f_{1T}^{\perp q}(\xbj,k_T^2) \frac{ ({\bm {\hat P}}
\times {\bm k}_T)\cdot {\bm S}}{M},
\label{eq:sivers}
\ee
where ${\bm S}$ is the spin of the target nucleon and
${\bm {\hat P}}$ is a unit vector opposite to the direction of the
virtual photon momentum. The fact that such a term
may be present in (\ref{eq:sivers}) is known as the Sivers effect
and the function $f_{1T}^{\perp q}(\xbj,k_T^2)$
is known as the Sivers function.
The latter vanishes in a naive parton 
picture since $({\bm {\hat P}} \times {\bm k}_T)\cdot {\bm S}$ 
is odd under naive time reversal (a property known as naive-T-odd), 
where one merely reverses
the direction of all momenta and spins without interchanging the
initial and final states. The momentum fraction $x$, 
which is equal to $\xbj$ in DIS experiments, 
represents the longitudinal momentum of the quark {\it before}
it absorbs the virtual photon, as it is determined solely from the 
kinematic properties of the virtual photon and the target nucleon. 
In contradistinction, the transverse momentum ${\bm k}_T$ is 
defined in terms of the kinematics
of the final state and hence it represents the asymptotic 
transverse momentum
of the active quark {\it after} it has left the target and before it
fragments into hadrons. Thus the Sivers function for semi-inclusive
DIS includes the final state interaction 
between struck quark and target remnant, and
time reversal invariance no longer requires that it vanishes.
\begin{figure}
\unitlength1.cm
\begin{picture}(10,2.5)(2.8,14)
\includegraphics{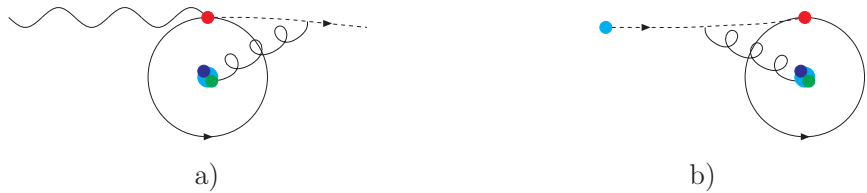}
\end{picture}
\caption{In SIDIS (a) the ejected (red) quark is attracted by
the (anti-red) spectators. In contradistinction, in DY (b), 
before annihilating with the (red) active quark, the approaching
(anti-red) antiquark is repelled by the (anti-red) spectators.}
\label{fig:SIDISDY}
\end{figure}
Indeed, as time reversal not only reverses the signs of all
spins and momenta, but also transforms final state interactions (FSI)
into initial state interactions (ISI), 
it has been shown that the Sivers 
function relevant for SIDIS and that relevant for 
Drell-Yan (DY) processes must have opposite signs 
[\refcite{collins}],
\be
f_{1T}^\perp(\xbj,{k}_T^2)_{SIDIS} =
- f_{1T}^\perp(\xbj,{k}_T^2)_{DY} ,
\label{SIDISDY}
\ee
where the asymmetry in DY arises from the ISI between the 
incoming antiquark and the target.
The experimental verification of this relation 
would provide a test of the current understanding of the Sivers 
effect within QCD. 
It is instructive to elucidate its physical
origin in the context of a perturbative picture:
for instance, when the virtual photon in a DIS process hits a red 
quark, the spectators must be collectively anti-red in order to
form a color-neutral bound state, and thus attract
the struck quark (Fig. \ref{fig:SIDISDY}). 
In DY, when an anti-red antiquark annihilates with
a target quark, the target quark must be red in order to merge
into a photon, which carries no color. Since the proton was 
colorless before the scattering, the spectators must be anti-red
and thus repel the approaching antiquark.

\begin{figure}
\unitlength1.cm
\begin{picture}(10,2.3)(3.,19.2)
\includegraphics{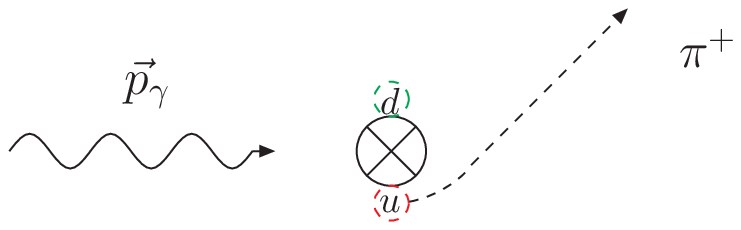}
\end{picture}
\caption{The transverse distortion of the parton cloud for a proton
that is polarized into the plane, in combination with attractive
FSI, gives rise to a Sivers effect for $u$ ($d$) quarks with a
$\perp$ momentum that is on the average up (down).}
\label{fig:deflect}
\end{figure}

The significant distortion of parton distributions in impact 
parameter space (\ref{eq:deform})
provides a natural mechanism for a Sivers effect.
In semi-inclusive DIS, when the 
virtual photon strikes a $u$ quark in a $\perp$ polarized proton,
the $u$ quark distribution is enhanced on the left side of the target
(for a proton with spin pointing up when viewed from the virtual 
photon perspective). Although in general the final state 
interaction (FSI) is very complicated, we expect it to be on average attractive thus translating a position space
distortion to the left into a momentum space asymmetry to the right
and vice versa (Fig. \ref{fig:deflect}) [\refcite{mb:SSA}].
Since this picture is very intuitive, a few words
of caution are in order. First of all, such a reasoning is strictly 
valid only in mean field models for the FSI as well as in simple
spectator models [\refcite{spectator}]. 
Furthermore, even in such mean field models
there is no one-to-one correspondence between quark distributions
in impact parameter space and unintegrated parton densities
(e.g. Sivers function). While both are connected by a Wigner
distribution [\refcite{wigner}], 
they are not Fourier transforms of each other.
Nevertheless, since the primordial momentum distribution of the quarks
(without FSI) must be symmetric we find a qualitative connection
between the primordial position space asymmetry and the
momentum space asymmetry (with FSI). 
Another issue concerns the $x$-dependence of the Sivers function.
The $x$-dependence of the position space asymmetry is described
by the GPD $E(x,0,-{\Delta}_\perp^2)$. Therefore, within the above
mechanism, the $x$ dependence of the Sivers function should be
related to the $x$ dependence of $E(x,0,-{\Delta}_\perp^2)$.
However, the $x$ dependence of $E$ is not known yet and we only
know the Pauli form factor $F_2=\int {\rm d}x E$. Nevertheless, 
if one makes
the additional assumption that $E$ does not fluctuate as a function 
of $x$ then the contribution from each quark flavor $q$ to the
anomalous magnetic moment $\kappa$ determines the sign of 
$E^q(x,0,0)$
and hence of the Sivers function. Making these assumptions,
as well as the very plausible assumption that the FSI is on average
attractive, 
one finds that $f_{1T}^{\perp u}<0$, while 
$f_{1T}^{\perp d}>0$. Both signs have been confirmed by a flavor
analysis based on pions produced in a SIDIS experiment 
by the {\sc Hermes} collaboration [\refcite{hermes}] and are
consistent with a vanishing isoscalar Sivers function
[\refcite{compass}].

\section{The Force on a Quark in SIDIS}
The chirally even spin-dependent twist-3 parton distribution 
$g_2(x)=g_T(x)-g_1(x)$ is defined as
\begin{eqnarray}
& &\int \frac{d\lambda}{2\pi}e^{i\lambda x}
\langle PS|\bar{\psi}(0)\gamma^\mu\gamma_5\psi(\lambda n)
|_{Q^2}|PS\rangle
\nonumber\\
& &\qquad=
2\left[g_1(x,Q^2)p^\mu (S\cdot n) 
+ g_T(x,Q^2)S_\perp^\mu +M^2 g_3(x,Q^2)n^\mu (S\cdot n)
\right].
\nonumber
\end{eqnarray}
neglecting $m_q$: $g_2(x)=g_2^{WW}(x)+\bar{g}_2(x)$, with
$g_2^{WW}(x)=-g_1(x)+\int_x^1 \frac{dy}{y}g_1(y)$.
$\bar{g}_2(x)$ involves quark-gluon correlations, e.g.
[\refcite{Shuryak,Jaffe}]
\be
\int dx x^2 \bar{g}_2(x)= \frac{d_2}{6}
\ee
with
\be
g\left\langle P,S \left|\bar{q}(0)G^{+y}(0)\gamma^+q(0) 
\right|P,S\right\rangle =
M {P^+}P^+ S^x d_2
\label{eq:twist3}
\ee
At low $Q^2$, $g_2$ has the physical interpretation of a spin 
polarizability, which is why the matrix elements (note that
$\sqrt{2}G^{+y}=B^x-E^y$) 
\be
\chi_E 2M^2 {\vec S} = \left\langle P,S\right|
q^\dagger {\vec \alpha} \times g {\vec E} q \left| P,S\right\rangle
\quad\quad
\chi_B 2M^2 {\vec S} = \left\langle P,S\right|
q^\dagger g {\vec B} q \left| P,S\right\rangle
\ee
are sometimes called spin polarizabilities or color electric and 
magnetic polarizabilities [\refcite{Ji}]. In the following we will 
discuss that at
high $Q^2$ a better interpretation for these matrix elements is that
of a `force'.

As Qiu and Sterman have shown [\refcite{QS}], the average transverse
momentum of the ejected quark (here also averaged over the momentum
fraction $x$ carried by the active quark)
in a SIDIS experiment can be represented by the matrix element
\be
\langle k_\perp^y\rangle = - \frac{1}{2P^+}
\left\langle P,S \left|\bar{q}(0)\int_0^\infty dx^-G^{+y}(x^+=0,x^-)
\gamma^+q(0) \right|P,S\right\rangle
\label{eq:QS}
\ee
which has a simple physical interpretation: the average transverse 
momentum is obtained by integrating the transverse component
of the color Lorentz force along the trajectory of the active quark
--- which is an almost light-like trajectory along the 
$-\hat{z}$ direction, with $z=-t$: 
The $\hat{y}$-component of the Lorentz force acting on 
a particle moving, with (nearly) the speed of light
${\vec v}=(0,0,-1)$, along the $-\hat{z}$ direction reads
\be
g\sqrt{2}G^{y+} = g\left(E^y + B^x\right) = g\left[ {\vec E} + {\vec v}\times
{\vec B}\right]^y.
\ee

We now rewrite Eq. (\ref{eq:QS}) as an integral over time
\be
\langle k_\perp^y\rangle = - \frac{\sqrt{2}}{2P^+}
\left\langle P,S \right|\bar{q}(0)\int_0^\infty dt G^{+y}(t,z=-t)
\gamma^+q(0) \left|P,S\right\rangle
\label{eq:QS2}
\ee
in which the physical interpretation of $- \frac{\sqrt{2}}{2P^+}
\left\langle P,S \right|\bar{q}(0)G^{+y}(t,z=-t)
\gamma^+q(0) \left|P,S\right\rangle$ as being the averaged force 
acting on the struck quark at time $t$ after being struck by the
virtual photon becomes more apparent. 

In particular,
\be
\label{eq:QS3}
F^y(0)&\equiv& - \frac{\sqrt{2}}{2P^+}
\left\langle P,S \right|\bar{q}(0) G^{+y}(0)
\gamma^+q(0) \left|P,S\right\rangle\\
&=& -\frac{1}{\sqrt{2}} MP^+S^xd_2
= -\frac{M^2}{2}d_2,
\nonumber
\ee
where the last equality holds only in the rest frame 
($p^+=\frac{1}{\sqrt{2}}M$) and for $S^x=1$,
can be interpreted as the averaged transverse
force acting on the active quark
in the instant right after it has been struck by the virtual photon.

Lattice calculations of the twist-3 matrix element yield 
[\refcite{latticed2}]
\be
d_2^{(u)} = 0.010 \pm 0.012 \quad \quad \quad \quad
d_2^{(d)} = -0.0056 \pm 0.0050
\ee
renormalized at a scale of $Q^2=5$ GeV$^2$ for the smallest
lattice spacing in Ref. [\refcite{latticed2}].
Here the identity $M^2\approx 5$GeV/fm is useful
to better visualize the magnitude of the force.
\be
F_{(u)} = -25 \pm 30 {\rm MeV/fm}\quad \quad \quad \quad
F_{(d)} = 14 \pm 13 {\rm MeV/fm}.
\ee
In the chromodynamic lensing picture, one would have expected
that $F_{(u)}$ and $F_{(d)}$ are of about the same magnitude and with
opposite sign. The same holds in the large $N_C$ limit.
A vanishing Sivers effect for an isoscalar target would be more
consistent with equal and opposite average forces. However, since
the error bars for $d_2$ include only statistical errors, the 
lattice result may not be inconsistent with 
$d_2^{(d)} \sim - d_2^{(u)}$.

The average transverse momentum from the Sivers effect is
obtained by integrating the transverse force to infinity
(along a light-like trajectory) 
$\langle k^y\rangle = \int_0^\infty dt F^y(t)$. This motivates us to
define an `effective range' 
\be
R_{eff} \equiv \frac{\langle k^y\rangle}{F^y(0)}.
\ee
Note that $R_{eff}$ depends on how rapidly the correlations fall
off along a light-like direction and it may thus be larger than
the (spacelike) radius of a hadron. 
Of cource, unless the functional form of the integrand is known,
$R_{eff}$ cannot really tell us about the range of the FSI,
but if the integrand does not oscillate

Fits of the Sivers function 
to SIDIS data yield
\refcite{Mauro} one finds about $|\langle k^y\rangle|\sim
100$ MeV [\refcite{Mauro}]. 
Together with the (average) value for $|d_2|$ from
the littice this translates into an effective range $R_{eff}$ of 
several fm.
It would be interesting to compare $R_{eff}$ for different quark 
flavors and as a function of $Q^2$, but this requires more
precise values for $d_2$ as well as the Sivers function.

Note that a complementary approach to the effective range was 
chosen in Ref. \refcite{Stein}, where the twist-3 matrix element
appearing in Eq. (\ref{eq:QS3}) was, due to the lack of lattice 
QCD results, estimated using QCD sum rule techniques. Moreover,
the `range' was taken as a model {\em input} parameter to estimate
the magnitude of the Sivers function.

A measurement of the twist-4 contribution $f_2$
to polarized DIS
allows determination of the expectation value of different 
Lorentz/Dirac components of the quark-gluon correlator appearing
in (\ref{eq:twist3})
\be
f_2 M^2S^\mu = \frac{1}{2} \left\langle p,S\right|
\bar{q}g\tilde{G}^{\mu \nu}\gamma_\nu q\left|p,S\right\rangle , 
\ee
In combination with (\ref{eq:twist3}) this 
allows a decomposition of the force into electric and magnetic
components using
\be
F_E^y(0)= - \frac{M^2}{8} \chi_E\quad \quad \quad \quad
F_B^y(0)= - \frac{M^2}{4} \chi_B
\ee
for a target nucleon polarized in the $+\hat{x}$ direction, where
[\refcite{Ji,color}]
\be
\chi_E = \frac{2}{3}\left(2d_2+f_2\right)
\quad \quad \quad \quad
\chi_M = \frac{1}{3}\left(4d_2-f_2\right).
\ee

A relation similar to (\ref{eq:QS3}) can be derived for the
$x^2$ moment of the twist-3 scalar PDF $e(x)$. For its
interaction dependent twist-3 part $\bar{e}(x)$ one finds for an
unpolarized target [\refcite{Yuji}] 
\be
4MP^+P^+ e_2 &=& 
g\left\langle p\right|\bar{q}\sigma^{+i}G^{+i}q
\left|P\right\rangle,
\label{eq:odd1}
\ee
where $e_2\equiv \int_0^1 dx x^2\bar{e}(x)$.
The matrix element on the r.h.s. of Eq. (\ref{eq:odd1})
can be related to the average transverse force acting on
a transversely polarized quark in an unpolarized target right after 
being struck by the virtual photon. Indeed, for the
average transverse momentum in the $+\hat{y}$ direction,
for a quark polarized in the $+\hat{x}$ direction, one finds
\be
\langle k^y \rangle = \frac{1}{4P^+}\int_0^\infty dx^-
g\left\langle p\right| \bar{q}(0) \sigma^{+y}G^{+y}(x^-)q(0)\left|
p\right\rangle
\label{eq:odd2}.
\ee
A comparison with Eq. (\ref{eq:odd1}) shows that the average 
transverse force at $t=0$ (right after being struck) on a
quark polarized in the $+\hat{x}$ direction reads
\be
F^y(0) = \frac{1}{2\sqrt{2}p^+} g\left\langle p\right| 
\bar{q} \sigma^{+y}G^{+y}q\left|
p\right\rangle = \frac{1}{\sqrt{2}}MP^+S^x e_2 = \frac{M^2}{2} e_2,
\ee
where the last identify holds only in the rest frame of the target
nucleon and for $S^x=1$. 

The impact parameter distribution for quarks polarized in
the $+\hat{x}$ direction was found to be shifted in the
$+\hat{y}$ direction [\refcite{DH,latticeBM,hannafious}].
Applying the chromodynamic lensing model implies a force
in the negative $-\hat{y}$ direction for these quarks and one
thus expects $e_2<0$ for both $u$ and $d$ quarks. 
Magnitude: since 
$\kappa_\perp>\kappa$, expect odd force larger than even force
and thus $|e_2| > |d_2|$.

It would be interesting to study not only whether the effective range
is flavor dependent, but also whether there is a difference between 
the chirally even and odd cases. It would also be very interesting
to learn more about the time dependence of the FSI by calculating
matrix elements of $\bar{q}\gamma^+ \left(\partial^+ G^{+\perp}
\right)q$, or even higher derivatives, in lattice QCD.
Knowledge of not only the value of the integrand at the origin,
but also its slope and curvature at that point, would be very
useful for estimating the integral in Eq. (\ref{eq:QS}).

\section*{Acknowledgments} I would like to thank the organizers of
the Transversity 2008 workshop for the kind invitation and
Daniel Boer and
Yuji Koike for very helpful discussions. This work was supported
in part by the DOE under grant number DE-FG03-95ER40965.

\bibliographystyle{ws-procs9x6}

\end{document}